\documentclass[showpacs,twocolumn,pre]{revtex4-1}
\usepackage{amssymb,amsmath,amsthm,mathtools,color,array}
\newcolumntype{P}[1]{>{\centering\arraybackslash}p{#1}}

\newcommand{\derpar}[2] {\frac{\partial #1}{\partial #2}}

\begin{document}
\title{Universality of active wetting transitions}

\author{N\'estor Sep\'ulveda}
\author{Rodrigo Soto}
\affiliation{Departamento de F\'{\i}sica, Facultad de Ciencias F\'{\i}sicas y Matem\'aticas, Universidad de Chile, Avenida Blanco Encalada 2008, Santiago, Chile}

\date{\today}

\begin{abstract}
Four on-lattice and six off-lattice models for active matter are  studied numerically, showing that in contact with a wall, they  display universal wetting transitions between three distinctive phases. 
The particles, which interact via exclusion volume only, move persistently and, depending on the model,  change their direction either via tumble processes or rotational diffusion. 
When increasing the turning rate $\nu_T$, the systems transit from total wetting, to partial wetting and dewetted phases. In the first phase, a wetting film covers the wall, with increasing heights while decreasing $\nu_T$. The second phase is characterized by wetting droplets on the walls. And, finally, the walls dries with few particles in contact with it.
These phases present two continuous non-equilibrium transitions. For the first transition, from partial to total wetting, the fraction of dry sites vanishes continuously when decreasing $\nu_T$, with a power law of exponent 1. And, for the second transition, an order parameter proportional to the excess mass in droplets
decreases continuously with a power law of exponent 3 when approaching a critical value of $\nu_T$. The critical exponents are the same for all the models studied.
\end{abstract}
\maketitle

\section{ Introduction}

Self-propelled active particles are a well established model for active matter~\cite{Vicsek2012,Marchetti2013,ABP,ABPRTP}. Particles move at constant speed with a director that changes randomly due to Brownian-like rotational diffusion or due to random tumble events. Individually, they present an effective spatial diffusivity much larger than the Brownian one and can replicate chemotaxis in presence of external chemical gradients~\cite{BergBrownian,BergEColi}. When interactions between particles are connected, new phenomena emerge with swarming, clustering, and phase separation being the most studied~\cite{Vicsek2012,Marchetti2013,cates2015}. Also, collective effects near interfaces have been reported~\cite{Wensink2008,Costanzo2012,Figeroa2015}.
In presence of an impenetrable wall, the persistence in motion gives rise to an accumulation of particles on the walls, in an ``active wetting'' phenomenon.
In Ref.~\cite{SS2017}, considering lattice models, we showed that active wetting develops some properties similar to those at equilibrium: total wetting, partial wetting, and dewetted phases were characterized, with transitions that present critical properties.

Molecular systems at equilibrium, interacting with solid surfaces, can present several wetting states, depending on the temperature and surface tension~\cite{deGennes}. In particular, a transitions between phases of partial wetting, with finite contact angle, and total wetting, where the contact angle vanishes strictly and thick films can form, takes place close to the bulk critical point~\cite{Cahn,Moldover}. This equilibrium wetting transition presents universal critical properties~\cite{Nakanishi,deGennes}.
The universality of the critical exponents in equilibrium transitions is a well stablished fact, well accounted by the renormalization group theory~\cite{Stanley,Zinn,Binney}. Though, in out-of-equilibrium conditions, as it is the case for active matter, the situation is less clear. While important contributions have been made using the dynamic renormalization group (see  \cite{Hohenberg77} and the recent review \cite{Tauber2014}) or in the seminal work of Toner and Tu for active flocks \cite{Toner1998},  still much research is needed. This article aims to contribute to this understanding using active matter as a prototype of non-equilibrium systems, analyzing the active wetting transitions.

Considering four on-lattice  and six off-lattice models of self-propelled active particles, moving in two dimensions (2D),  we aim to determine the universal character of the active wetting transitions in presence of a non attractive wall.
In these models, the interaction between particles is only repulsive to describe excluded volume, and hydrodynamics interaction or torques induced by other particles or by the wall are not considered. Partial or total overlapping is allowed, to account for active particles that can deform or move partially in a third dimension, as it is the case of sedimenting colloids. Particles change direction at a turning rate $\nu_T$ via rotational diffusion or tumble processes, to model the dynamics of different kind self-propelled agents displaying persistent motion, for example, swimming bacteria, migrating cell in cultures, active Janus colloids, among others~\cite{bechinger2016}.

Appropriate order parameters are considered and the critical exponents are obtained resulting in a test for the universality of the transitions. 
 In the  on-lattice models considered in this work, particles move in a regular lattice at discrete time steps with directions that change randomly at a small tumbling rate $\alpha$. Hence, the scale for the turning rate is $\nu_T=\alpha$.
Excluded volume is achieved by imposing a nominal maximum occupation per site $n_\text{max}$, which reproduce particle overlaps: if $n_\text{max}=1$  the interaction between particles is steric, while larger values allow for increasing number of particles that overlap. 
In the off-lattice case, particles move in continuous time and space. The swimming directions can change either via tumble processes, with a rate $\alpha$, or via continuous rotational diffusion, characterized by a coefficient $D_r$, resulting in scales for the turning rate  that go as $\nu_T=\alpha$ or $\nu_T=D_r$, respectively.
The interaction between particles is repulsive only, and we consider either the  Weeks-Chandler-Andersen (WCA) or a Gaussian potential, both of strength $\epsilon$. 

For unbounded systems in absence of walls, depending on the turning rate and the number of particles that are allowed to overlap, particles can either aggregate in few big clusters after a coarsening process,  develop many  small clusters, or form an homogeneous gas, as it has been seen for lattice~\cite{thompson2011,soto2014,SS2016,slowman2016} and off-lattice cases~\cite{peruani2006,fily2012,therkauff2012,bialke2013,redner2013,buttinoti2013,palacci2013,levis2014,bialke2015,locatelli2015}.
Here, we show that in all models, for parameters corresponding to the bulk homogeneous gas phase, in presence of a wall, three phases can develop. At small $\nu_T$, the system presents total wetting where a thick film forms. Increasing   $\nu_T$, a transition to partial wetting takes place. Here, droplets dispose on the wall, decreasing their height when increasing  $\nu_T$. Finally, at higher turning rates, the wall dewets. These two transitions present critical behavior, with order parameters following power laws with exponents taking the same values for the ten models. 

The manuscript is organized as follows. In Sec.~\ref{sec.lattice} and \ref{sec.offlattice}, we describe, respectively, the on-lattice and the off-lattice models. Section~\ref{sec.transitions} presents the order parameters used to characterize the wetting transitions and obtain the associated exponents, which are shown to be universal. Finally, conclusions are given in Sec.~\ref{sec.conclusion}.

\section{On-lattice models} \label{sec.lattice}
We consider four variations of the Persistent Exclusion Process (PEP) described in Refs.~\cite{soto2014,SS2016,SS2017}, which models in a lattice the dynamics of active particles presenting run-and-tumble dynamics. A total of $N$ particles move on a regular 2D lattice. Each particle has a state variable $\mathbf{s}_k$ ($|\mathbf{s}_k|=1$, where $k$ denotes the particle),  indicating the direction which it points to. Time evolves at discrete steps in which particles attempt to jump one site in the direction pointed by $\mathbf{s}_k$. Depending on the occupation of the destination site, the jump is performed, otherwise the particle remains in the original position. The maximum occupancy $n_\text{max}$ can be larger than one to descrive active particles that can overlap. The position update of the particles is asynchronous to avoid two particles attempting to jump to the same site simultaneously:  at each time step, the $N$ particles are sorted randomly and, sequentially, each particle attempts to jump. 
After the position updates, with some small probability $\alpha$, independently for each particle, tumbles take place and the $\mathbf{s}_k$ changes to a new random direction~\cite{commentalpha}.
The four variations of the model differ in the lattice geometry, in the form the maximum occupancy is enforced, and the way the new directions are chosen after a tumble. 

\subsection{Model 1}
This case corresponds to the original PEP model. Here, the lattice is square, composed of $L_x\times L_y$ sites and, consistently, $\mathbf{s}_k =\{\pm \hat{\mathbf{x}}, \pm \hat{\mathbf{y}}\}$ (see Fig.~\ref{fig.models}a).
If the occupation of the destination site is smaller than the maximum occupancy per site $n_\text{max}$, the jump takes place; otherwise, the particle remains at the original position.  That is, the maximum occupancy is strictly enforced. 
For tumbles, the directors ${\mathbf s}_k$ are redrawn at random from the four possibilities, independently of the original value.
 
 \subsection{Model 2}
 Here, the lattice and the tumble protocol are the same as in Model 1. To describe swimmers or agents that can deform or be compressed, the jumps of the particles to the new site are now probabilistic, depending on occupation fraction of the destination site $n_{\mathbf{r}_k+\mathbf{s}_k}/n_\text{max}$,  where $\mathbf{r}_k$ is the position of the $k$-th particle. With probability
 \begin{equation}
 P_{\mathbf{r}_k\rightarrow \mathbf{r}_k+\mathbf{s}_k}=\exp\left[-\left(\frac{n_{\mathbf{r}_k+\mathbf{s}_k}}{n_\text{max}}\right)^6 \right]
 \end{equation}
the jump is performed (see Refs.~\cite{SS2016,SS2017}). As a result, sites can have higher occupancies than $n_\text{max}$, although with low probability. The sixth power has been chosen such that the allowed occupancies of a sites do not exceed for too much the specified value $n_\text{max}$. Larger powers produce jump probabilities that are too sharp and the model is equivalent to having integer $n_\text{max}$. For smaller powers, the jump probability is extremely smooth and actual occupations can exceed notoriously $n_\text{max}$. Short tests made with other exponents close to 6 give similar qualitative results.
  
 \subsection{Model 3}
 In this case, the lattice and the maximum occupancy enforcement are as in Model 1. The modification here is associated to the change of direction, to make it similar to the case of rotational diffusion. When a tumble takes place, the new director $\mathbf{s}_k$ is sorted at random between the two directions perpendicular to the pre-tumble director, introducing a short time memory in the dynamics. 
 
  \subsection{Model 4}
  
\begin{figure}
\includegraphics[width=\columnwidth]{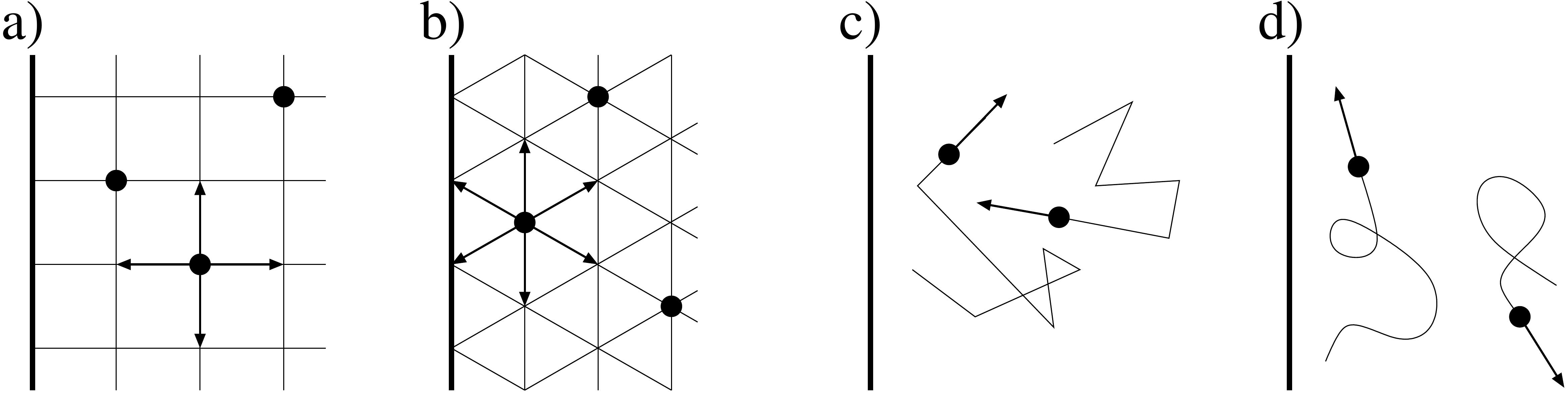}
\caption{a) Lattice used for models 1, 2, and 3. Particles can move in any of the four directions indicated by the arrows. b) Lattice used for model 4. Here, six directions are possible. c) RTP model, where particles move in straight lines between tumble events. d) Particles in the ABP model change direction continuously. In all cases, the wall is designed by a thick vertical line.}
\label{fig.models}
\end{figure}

Finally,  the lattice is changed to a triangular one  with a major axis parallel to the wall  (see Fig.~\ref{fig.models}b).  The six values of the state variable are $\mathbf{s}_k =\{\pm \hat{\mathbf{y}}, \pm \frac{\sqrt{3}}{2}\hat{\mathbf{x}} \pm \frac{1}{2}\hat{\mathbf{y}}\}$. Maximum occupancy is enforced as in Model 1 and the tumbles have no memory, but now the new direction is chosen with equal probability between the six directions.
 
 \section{Off-lattice models} \label{sec.offlattice}
 For the off-lattice cases we consider the Run and Tumble Particle (RTP) and the Active Brownian Particle (ABP)  models of active particles~\cite{ABP,ABPRTP}.
 A total of $N$ particles move in a continuous 2D surface of area $L_x\times L_y$ and time evolves continuously. 
Each particle is characterized by its position $\mathbf r_k$ and  director $\mathbf{s}_k=(\cos\theta_k,\sin\theta_k)$ , with  $\theta_k\in\mathbb{R}$. Both on- and off-lattice cases, $\mathbf{s}_k$ corresponds to a unitary vector and has an analogous meaning. 
The positions  evolve as
 \begin{equation}
\frac{{\rm d} \mathbf r_k}{{\rm d} t} = V_0 \mathbf{s}_k - \mu \derpar {}{\mathbf r_k} \sum_{{k'}\neq k} U(\mathbf r_k -\mathbf r_{k'}),
\end{equation}
where $V_0$ is the self-propulsion speed, $\mu$ is a mobility which is normally absorbed in the potential (thus, we take $\mu=1$ henceforth), and $U$ is a repulsive potential, either a WCA potential
\begin{equation}
U(\mathbf{r})=
\begin{cases}
4\epsilon\left[(\sigma/r)^{12}-(\sigma/r)^{6}\right]+\epsilon, &  \text{if  } r<2^{1/6}\sigma \\  
0,                                                                                                     &  \text{if  } r\geq2^{1/6}\sigma
\end{cases},
\end{equation}
or a Gaussian potential
\begin{equation}
U(\mathbf{r})=\epsilon\, e^{-(r/\sigma)^2}.
\end{equation}
Here $\epsilon$ is the strength of the potential and $\sigma$  is the diameter of the particles. As for the lattice cases, particles only interact via excluded volume effects. The main difference between the two potentials is that with WCA particles can never overlap completely (see first row of Fig.~\ref{fig.Model9} or Fig.~\ref{fig.Model10}), contrary to the Gaussian potential, where particles can occupy the same position (see first row of Fig.~\ref{fig.Model8}). The Gaussian potential is appropriate for geometries where particles can move partially in the third dimension and, hence, appear to overlap in the plane.

There are no explicit modification of the propulsion speed due to interactions, neither Vicsek-like alignment mechanisms for the directors. Indeed,  directors evolve independently for each particle. 
For the RTP model, with a rate $\alpha$,  particles can experience a tumble (see Fig.~\ref{fig.models}c). In an instantaneous event, the direction changes to a  new one $\mathbf{s}'_k$, which is chosen from a kernel $W(\mathbf{s}_k,\mathbf{s}'_k)$ that could depend on the original director. For  isotropic media  $W(\mathbf{s}_k,\mathbf{s}'_k)=w(\mathbf{s}_k\cdot\mathbf{s}'_k)$ and we take $W=1/2\pi$, indicating that the new director is chosen completely at random.
In the case of the ABP model (see Fig.~\ref{fig.models}d), the director $\mathbf{s}_k$ changes continuously as an effect of rotational noise, process that is described with a Langevin equation
 \begin{equation}
\frac{{\rm d} \mathbf{s}_k}{{\rm d}t} = \sqrt{2D_r}\, \mathbf{s}_k\times \pmb{\eta}_k(t), \label{eq.abp}
\end{equation}
where $D_r$ is an effective rotational diffusion coefficient and $\pmb{\eta}_k=\eta_k\mathbf{\hat{z}}$ ($\mathbf{\hat{z}}$ is perpendicular to the plane where the particles move) is a white noise of zero mean and correlation $\langle \eta_{k}(t) \eta_{k'}(t')\rangle = \delta_{kk'} \delta(t-t')$, where $k$ and $k'$ label the particles. Eq.~(\ref{eq.abp}) is equivalent to $\dot\theta_k=\sqrt{2D_r}\eta_k(t)$ in 2D.

The wall is modeled in three different ways. 
First, we consider rigid hard walls (Wall A), which we implement as follows:  after a time step, when a particle crosses the wall, it is displaced back, placed in contact with the wall, without changing the $y$ coordinate.
To model  swimmers that can deform in contact with solid surfaces, we consider walls that exert smooth repulsive forces in the $x$ direction (Wall B). It is implemented with a WCA repulsive potential, with the same value of $\epsilon$ as for the interparticle interactions.
Finally, the Wall C model is sticky, to described cases where swimmers cannot slide in contact with a solid surface. In this case, when the $x$-coordinate of the particle is at a distance $\sigma/2$ or smaller to the wall and $\hat{\mathbf{n}}\cdot\mathbf{s}\leq0$, with $\hat{\mathbf{n}}$ exterior normal of the wall, the particle velocity is set to zero. Otherwise, it is subject to a WCA repulsive potential as in Wall B case. Particles can escape from the sticky condition by changes in the director $\mathbf{s}$.
For all  models, no torque is induced by the wall.

    \subsection{Model 5}
 
 Here, particles interact with the WCA potential, the director evolves according to the RTP model, and  Wall A is used.
   
   \subsection{Model 6}
   
 The same as Model 5, but with directors evolving with the  ABP model.

 \subsection{Model 7}
 
The same as Model 5, but the Gaussian potential is used for the interaction between particles. 

 \subsection{Model 8}
 
Identical to  Model 6, but with the use of the Gaussian potential for the interaction between the particles.

 \subsection{Model 9}
Identical to  Model 6, but Wall B is used.
 
  \subsection{Model 10}
Identical to  Model 6, with the use of the sticky Wall C.

\smallskip

The equations of motion are integrated numerically using explicit Euler (Models 5 and 7) or Euler-Heun (Models 6, 8, 9 and 10) methods with a small time step (${\rm d}t=0.0005 \sigma/V_0$). Smaller time steps give equivalent results.

 \begin{figure*}
\includegraphics[width=2\columnwidth]{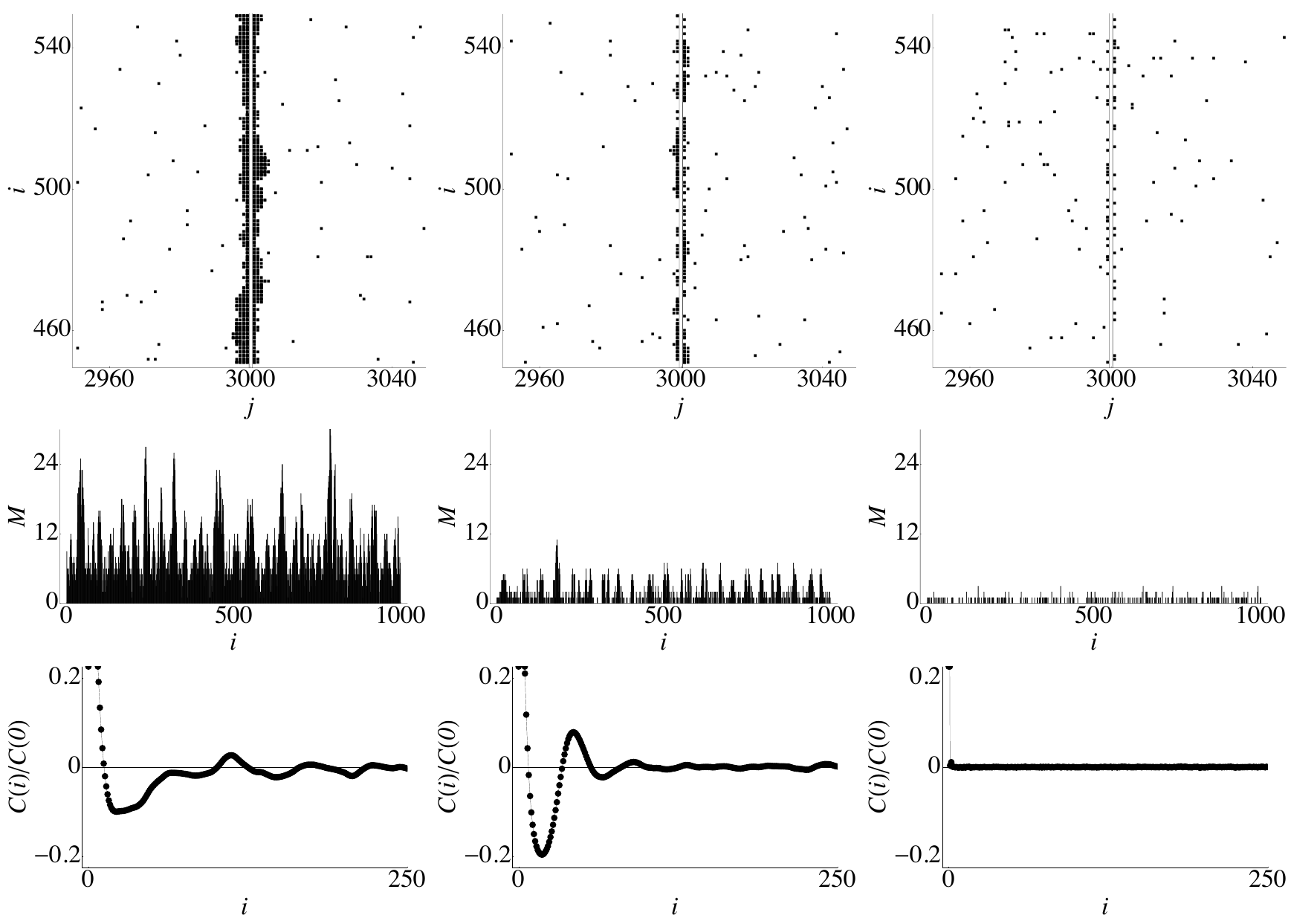}
\caption{
Model 2 with $n_\text{max}=3$, $\phi=0.01$, and increasing values of $\hat{\nu}_T=0.001, 0.005, 0.03$ (from left to right). 
First row: Snapshots of part of the system ($L_x=6000$ and $L_y=1000$). 
Second row: Instantaneous values of the mass profile $M(i)$. 
Third row: Normalized mass autocorrelation function $C(i)/C(0)$. 
}
\label{fig.Model2}
\end{figure*}

 \begin{figure*}
\includegraphics[width=2\columnwidth]{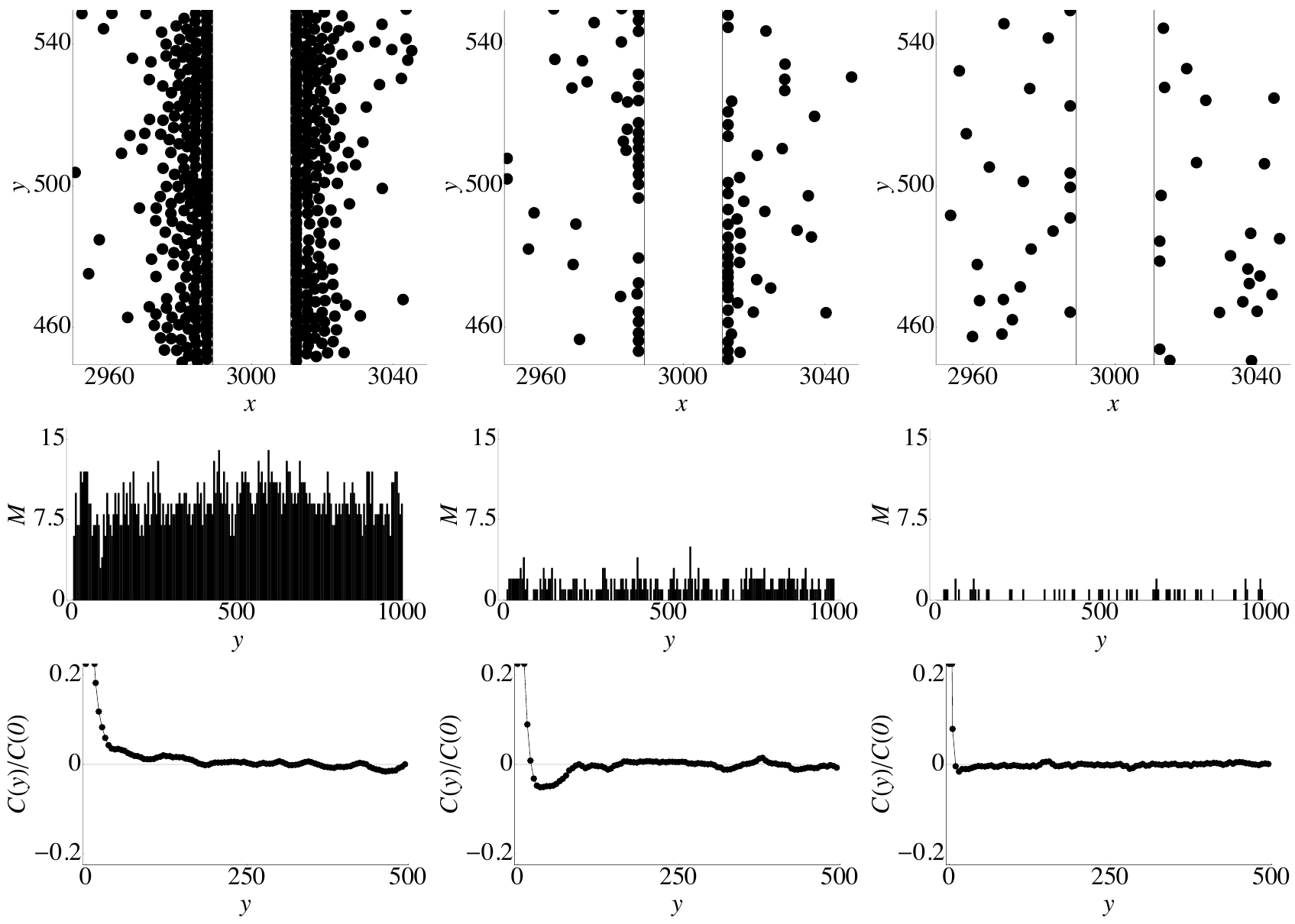}
\caption{
Model 8 with $\widehat{V}_0=0.2$, $\phi=0.01$, and increasing values of $\hat{\nu}_T=0.002,0.0152,0.1208$  (from left to right).
First row: Snapshots of part of the system ($L_x=6000$ and $L_y=1000$). 
Second row: Instantaneous values of the mass profile $M(y)$. 
Third row: Normalized mass autocorrelation function $C(y)/C(0)$. 
}
\label{fig.Model8}
\end{figure*}

 \begin{figure*}
\includegraphics[width=2\columnwidth]{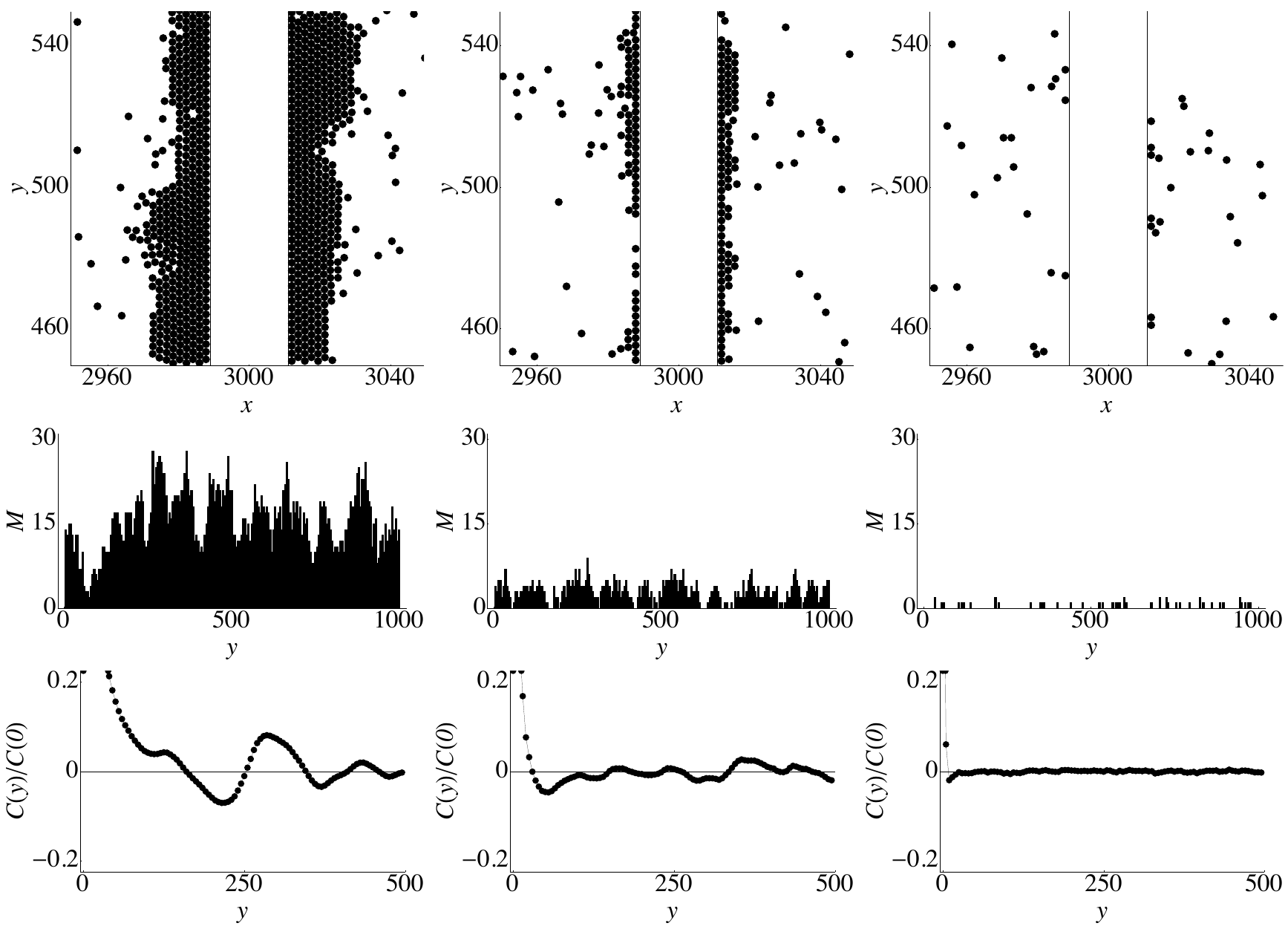}
\caption{
Model 9 with $\widehat{V}_0=2$, $\phi=0.01$, and increasing values of  $\hat{\nu}_T=0.002,0.0152,0.1208$  (from left to right).
First row: Snapshots of part of the system ($L_x=6000$ and $L_y=1000$). 
Second row: Instantaneous values of the mass profile $M(y)$. 
Third row: Normalized mass autocorrelation function $C(y)/C(0)$. 
}
\label{fig.Model9}
\end{figure*}

 \begin{figure*}
\includegraphics[width=2\columnwidth]{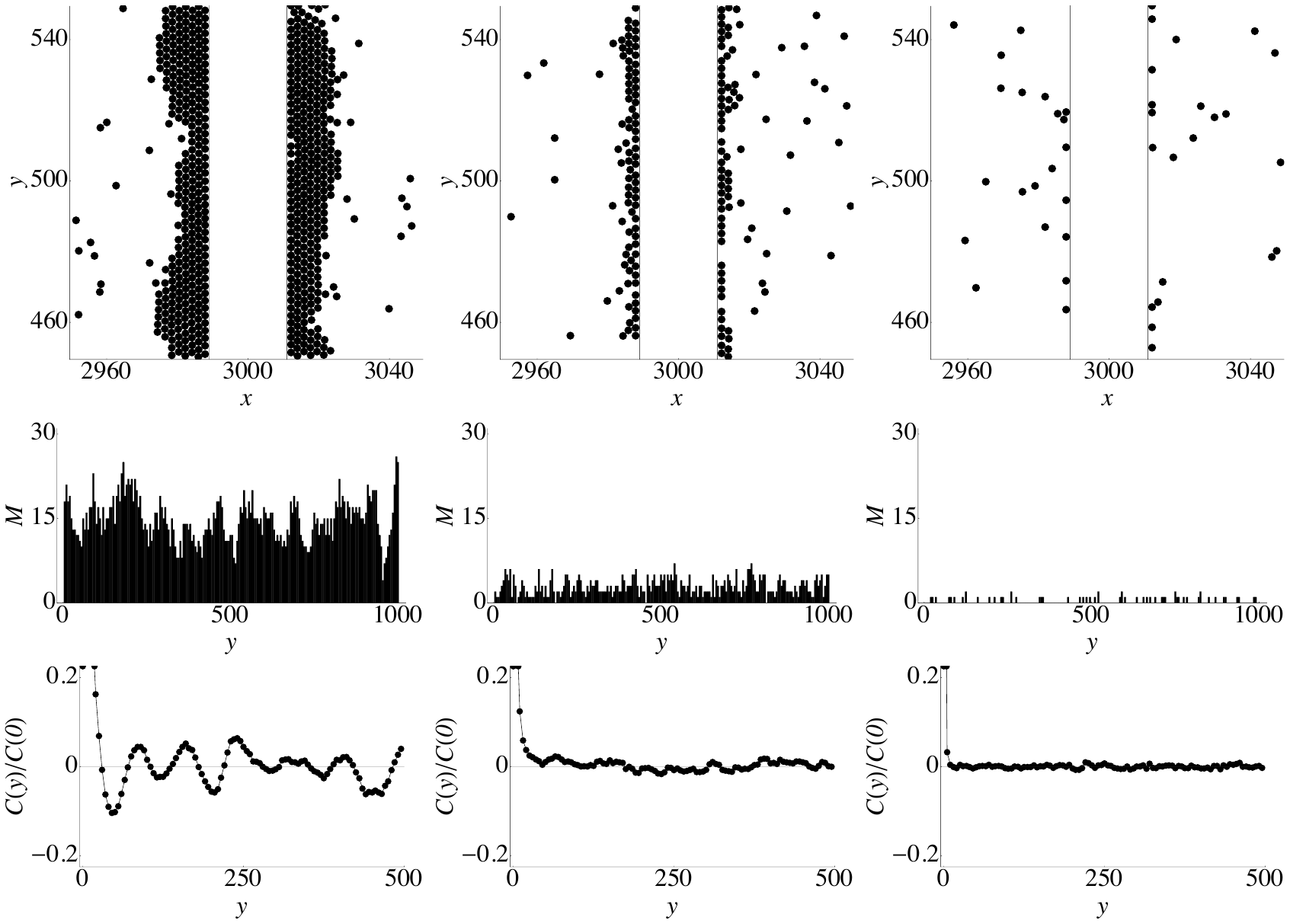}
\caption{
Model 10 with $\widehat{V}_0=2$, $\phi=0.01$, and increasing values of $\hat{\nu}_T=0.002,0.0086
,0.1208$ (from left to right).
First row: Snapshots of part of the system ($L_x=6000$ and $L_y=1000$). 
Second row: Instantaneous values of the mass profile $M(y)$. 
Third row: Normalized mass autocorrelation function $C(y)/C(0)$. 
}
\label{fig.Model10}
\end{figure*}

\section{Wetting transitions}\label{sec.transitions}
\subsection{Control parameters}

To study the active wetting phenomenon in the  models described in Sections \ref{sec.lattice} and \ref{sec.offlattice}, we consider a box with periodic boundary conditions and a vertical wall (parallel to the $y$ coordinate) placed in the middle of $L_x$. Particles pointing to the wall cannot reach it and can only move parallel or away from it, with the exception of  Model 10, where particles that point toward the wall can not slide.   
Particles that hit a wall will remain in contact with it for times comparable to $\nu_T^{-1}$. If this time is large, there is a high probability that during it, new particles arrive, blocking the first ones. The wall therefore, acts as a nucleation site for clusters and, naturally, wets by particles. The wetting film will increase by decreasing $\nu_T$ as has been observed under dilute conditions \cite{Elgeti2015,Ezhilan}. In Ref.~\cite{SS2017} it was shown that this tendency is not smooth and, rather, transitions take place for the lattice models.

The dimensionless control parameters for the on-lattice models are the global particle density $\phi=N/(L_x L_y)$, the turning rate $\hat{\nu}_T=\nu_T$, which is already dimensionless, and  the maximum nominal occupation number $n_\text{max}$. For the off-lattice models the control parameters are the area fraction $\phi=N\pi \sigma^2/(4L_x L_y)$, $\hat{\nu}_T=\nu_T \sigma/V_0$, and $\widehat V_0=V_0 \sigma /\epsilon$, which measures the degree of overlap that the particles can experience.
In part, the complexity of these models emerge because there are two dimensionless time scales: the aforementioned persistence time, $\hat\nu_T^{-1}$, and the mean flight time, which depends on density. 

Simulations are performed in  the  range of the parameters shown in Table~\ref{tablaParametros}. 
We have verified by simulations that in absence of walls, the systems remain in the gas phase for these parameters. The choice for the system sizes presented in Table~\ref{tablaParametros} has the following rationale~\cite{SS2017}. First, $L_y$ must be large enough to detect any spatial structuration of the wetting film that will be formed on the walls. Second, if $L_x$ is small, particles that evaporate from a wall after a change in  $\mathbf{s}_k$, could arrive to the other wall ballistically, creating artificial correlations between the two walls. 
At low concentrations and times longer than  $\nu_T^{-1}$ particles present an effective diffusive motion with a diffusion coefficient that scale as $D\sim 1/ \nu_T$ or $D\sim V_0^2/ \nu_T$ for the on-lattice and off-lattice models, respectively~\cite{BergBrownian}.  
To ensure that evaporated particles reach the diffusive regime, one must take $L_x\gg V_0\nu_T^{-1}$.  
For the initial condition, particles are placed randomly, respecting excluded volume, with random orientations. 
Finally, we consider total simulation times equal to $T=2\times 10^6$ and $T=4\times 10^4 \sigma/V_0$ for the on- and off-lattice case respectively, which are long enough to reach steady state and to achieve good statistical sampling. 
The persistent motion along the wall implies that even if the bulk density is small, the density in contact with the wall is large and particle--particle interactions become relevant.

\begin{table}
\caption{Values and ranges used for the parameters in the simulations of the ten models under study.}
\label{tablaParametros}
\begin{tabular}{P{0.8cm} P{1.8cm} P{1.2cm} P{1.4cm}  P{0.8cm} P{0.6cm} P{0.6cm} P{0.6cm}}  
\hline
\hline
Model & $\hat{\nu}_T$ & $n_\text{max}$ & $\widehat{V}_0$  & $\phi$ & $L_x$ & $L_y$ & Wall\\ 
\hline
\hline   
     1     & $[0.001,0.03]$   & $1, 2, 3, 4$  & $*$   &     0.010    & 6000 & 1000 & $*$\\ 
    2     & $[0.001,0.03]$   & $2, 3$   & $*$    &   0.010     & 6000 & 1000  & $*$\\ 
    3     & $[0.001,0.03]$   & $2, 3$   & $*$    &   0.010     & 6000 & 1000  & $*$\\ 
    4     & $[0.001,0.03]$   & $2, 3$   & $*$     &   0.040     & 3000 & 500 & $*$\\
    5     & $[0.002,0.20]$   & $*     $   & $0.0002, 2$      &  0.004  & 6000 &1000 & A\\
    6     & $[0.002,0.20]$ & $*     $     & $2$       &  0.004 & 6000 &1000 & A\\
    7     & $[0.002,0.20]$   & $*     $   & $0.2$      &  0.010  & 6000 &1000 & A\\
    8     & $[0.002,0.20]$ & $*     $     & $0.2$       &  0.010 & 6000 &1000 & A\\
    9     & $[0.002,0.20]$   & $*     $   & $0.0002, 2$      &  0.010  & 6000 &1000 & B\\
    10   & $[0.002,0.20]$ & $*     $     & $2$       &  0.010 & 6000 &1000 & C\\
\hline
\hline
\end{tabular}
\end{table}

\subsection{Order parameters}   
The wetting films are characterized as follows. First, for the analysis, we consider only wetting particles, defined as those that belong to clusters connected to the walls. For on-lattice models, clusters are made of particles belonging to nonempty contiguous sites  that have at least one site in contact with the wall. Once the connected particles are identified, for each vertical position $i$,  the associated mass, $M(i)$, corresponds to the total number of wetting particles in this row (see first row of Fig.~\ref{fig.Model2}). 
For off-lattice cases, the clusters are made of connected particles by distances smaller or equal than $\sigma$ for the interaction WCA and $1.6\sigma$ for the Gaussian interaction, that is they overlap at least partially, and at least one particle must be closer to the wall by a distance $\sigma/2$. The wall is divided in bins of length $2.5\sigma$ and the associated mass $M(i)$ is the total number of wetting particles with center in the bin (see first row of Figs.~\ref{fig.Model8}, \ref{fig.Model9}, and \ref{fig.Model10}).  This protocol used to compute $M(i)$  is not identical to that reported in Ref.~\cite{SS2017}, but we chose it because it gives less noisy results and it is well adapted for the on- and off-lattice models, while showing the same qualitative behavior and producing the same critical exponents.

By observing the instantaneous profiles of $M(i)$, three distinct phases are recognized when increasing $\hat \nu_T$ (see second row of Figs.~\ref{fig.Model2},  \ref{fig.Model8}, \ref{fig.Model9}, and \ref{fig.Model10}).
For small turning rates the mass is uniformly greater than one, corresponding to a phase of total wetting, where a continuous thick film forms. 
 Increasing the value of $\hat\nu_T$, droplets  appear, where regions of vanishing and finite values of $M(i)$ alternate, corresponding to partial wetting.  
 The emergence of droplets is a consequence of the particles motility and persistence. When a particle hits the wall it remains there or moves parallel to the wall until a change in $\mathbf{s}_k$ takes place. Then, if it hits another particle, it will block, acting then as a nucleation site for droplets. The balance between evaporation and the incoming flux of particles moving along the wall stabilizes the droplets size.
Note that, contrary to equilibrium wetting transitions, for all the studied models, droplets are thin with at most three particles in thickness.
Therefore it is not possible to define and compute a contact angle, neither it is possible to describe them using classical growth theories as the KPZ equation~\cite{KPZ}. Finally, for even larger turning rates, the thickness of the wetting layer is vanishingly small throughout the wall, which is interpreted as a dewetted  phase.

\begin{figure*}
\includegraphics[width=2.\columnwidth]{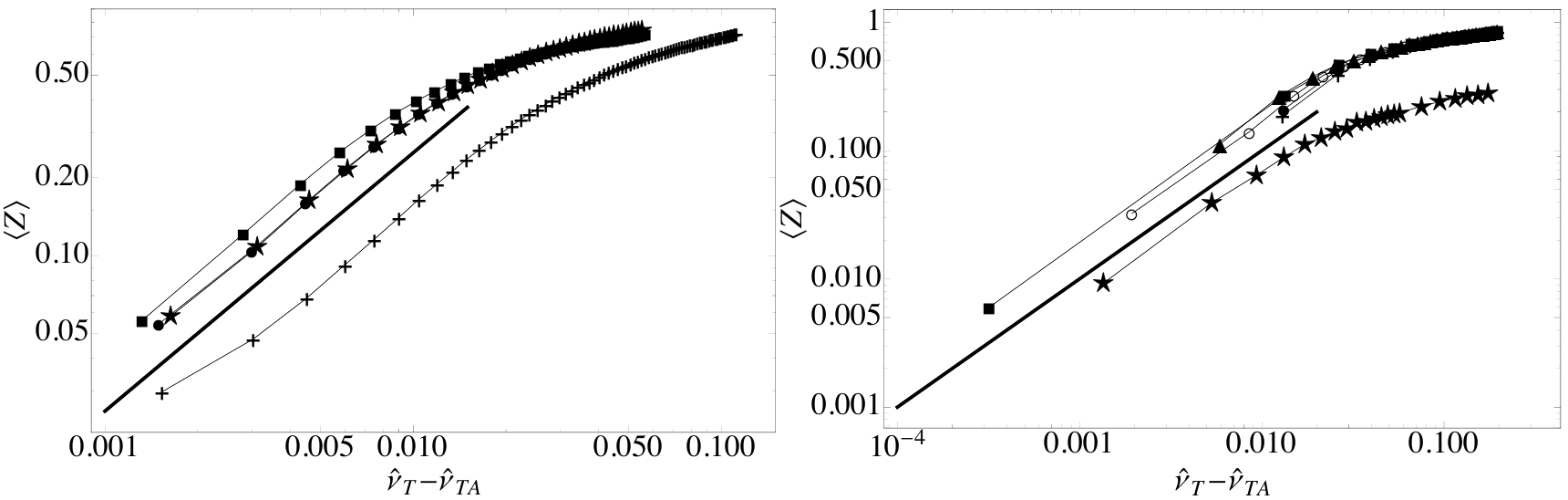}
\caption{
Critical behavior of the order parameter $Z$ above the total to partial wetting transition. The simulation results, with symbols, are joined by lines to help visualization. 
The thick solid lines present  power laws with exponent $1$.
Left: Models 1 (stars), 2 (squares), 3 (circles), 4 (crosses). Right: Models 5 (stars), 6 (squares), 7 (circles),  8 (crosses), 9 (triangles), and 10 (open circles).  
The area fractions are specified in Table~\ref{tablaParametros}, while $n_\text{max}=2$ for all the on-lattice models, $\widehat{V}_0=2$  for Models 5 and 6, $\widehat{V}_0=0.2$ for Models 7 and 8, and $\widehat{V}_0=2$  for Models 9 and 10.}
\label{fig.criticalZ}
\end{figure*}

To analyze the first transition, from total to partial wetting, we compute the instantaneous fraction of sites with vanishing mass, $Z=|{i : M(i)}=0|/L_y$. Its temporal average $\langle Z\rangle$ is an order parameter that vanishes when total wetting is achieved at a critical point $\nu_{TA}$ and presents a continuous transition (see Fig.~\ref{fig.criticalZ}). This transition from total to partial wetting, when $\hat{\nu}_T$ increases, is produced because evaporation becomes more important than the incoming flux of particles. 
In the on-lattice models, the  transition takes place in all cases, with the exception of $n_\text{max}=1$. In that case, $\langle Z \rangle$ never vanishes and rather decreases continuously, indicating that even when the wetting layer is thick, there are always a non-negligible fraction of empty sites. This results could be related to a previous observation that in bulk, clusters never coalesce and sizes are distributed exponentially for $n_\text{max}=1$~\cite{soto2014,SS2016}. Hence, the nucleation at the wall can not create a uniform thick film, even for small values of $\hat{\nu}_T$.
In the off-lattice cases, the transition always is observed except for very small values of $\widehat{V}_0$, where particles do not penetrate and the dynamics is analog to in-lattice models with $n_\text{max}=1$. Hence, some overlapping is needed to observe this transition. 
In the cases where the transition takes place, close to the critical point, $\langle Z\rangle$ follow  power laws, with exponents and critical turning rates that are obtained using the protocol described in Ref.~\cite{Castillo}. Within errors, the exponent is universal and all models are consistent with the law 
\begin{align}
\langle Z\rangle &= Z_0 \left(  \hat{\nu}_T - \hat{\nu}_{TA} \right)^{1.00\pm 0.04}, & \hat{\nu}_T&\geq \hat{\nu}_{TA},  \label{eq.fitZ}
\end{align}
as shown in Fig.~\ref{fig.criticalZ}. The fitted exponents and the non-universal  critical turning rates $\hat{\nu}_{TA}$ and  amplitudes $Z_0$  are given in the Table~\ref{tablaVC}. 

\begin{figure*}
\includegraphics[width=2.\columnwidth]{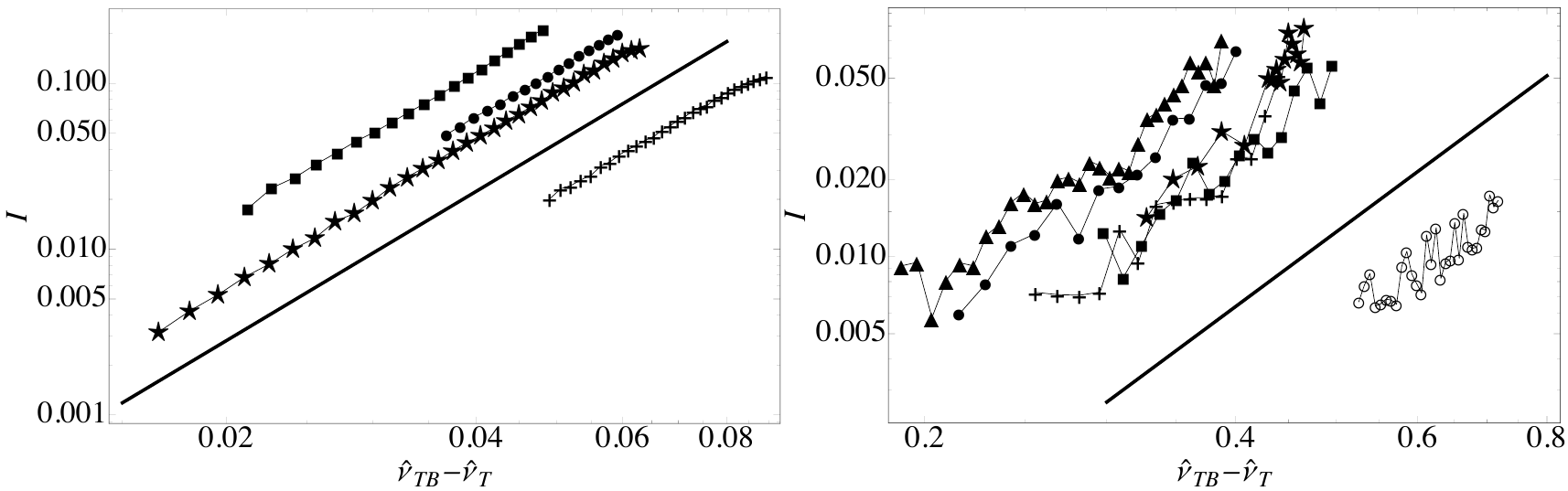}
\caption{
Critical behavior of the order parameter $I$ below the partial wetting to dewetting transition. The simulation results, with symbols, are joined by lines to help visualization.
The thick solid lines present  power law with exponent $3$.
Left: Models 1 (stars), 2 (squares), 3 (circles), 4 (crosses). Right: Models 5 (stars), 6 (squares), 7 (circles),  8 (crosses), 9 (triangles), and 10 (open circles).  
The area fractions are specified in Table~\ref{tablaParametros}, while  $n_\text{max}=3$ for the  on-lattice models, $\widehat{V}_0=2$   for Models 5 and 6, $\widehat{V}_0=0.2$  for Models 7 and 8, and $\widehat{V}_0=2$   for Models 9 and 10.}
\label{fig.criticalI}
\end{figure*}

To analyze the  structuration in droplets in the partial wetting to dewetting transition, we consider  the mass spatial autocorrelation function, $C(i) \equiv \langle M(i+i') M(i')\rangle -\langle M\rangle^2$.  
The average is done in the steady state over $i'$ and time. 
The third row of Figs.~\ref{fig.Model2}, \ref{fig.Model8}, \ref{fig.Model9}, and \ref{fig.Model10} display $C(i)/C(0)$ [$C(y)/C(0)$ for the off-lattice cases, with $y=2.5\sigma i$], for Models 2, 8, 9, and 10, respectively.
In the on-lattice cases, $C$ presents a maximum at a  distance $l$, which indicates that a periodic structure of droplets appears at intermediate values of $\hat\nu_T$. 
In the off-lattice models, the first maximum in the spatial autocorrelation function is absent: although droplets arrange with some regularity as can be seen in the plots of $M$ in Figs.~\ref{fig.Model9} and \ref{fig.Model10}, this order is not periodic. 
In Ref.~\cite{SS2017} we showed, for the four on-lattice models analyzed here, $P=\text{Prob}(M=n_\text{max})$ (corresponding to the probability that a specific row has $n_\text{max}$ particles in contact with the wall) is an order parameter for the transition, vanishing at the  critical tumbling rate. Here, we do not consider this parameter, because it is not pertinent for the off-lattice cases, where it is not possible to define a maximum occupancy per site.
Thus, to describe the partial wetting to dewetting transition we consider here an order parameter pertinent to all models: although there no maximum in $C(i)$, there is always a clear negative minimum with a depth $I$ (defined positive) that approaches zero when increasing $\hat\nu_T$ (see Figs.~\ref{fig.Model2}, \ref{fig.Model8}, \ref{fig.Model9}, and \ref{fig.Model10}, third row). 
The existence of a negative minimum in the autocorrelation function, for conditions where the number of empty sites $Z$ is finite, is a signature that there is an alternation in regions with and without particles, that is, the formation of droplets. If $\langle M \rangle$ is the average mass on all bins, the depth $I$ is a measure of the excess mass in the droplets compared to the average, $I\sim (M_\text{droplet}- \langle M \rangle) \langle M \rangle$.
The order parameter $I$ is not standard in the description of the partial wetting to dewetting transition in  equilibrium, but in this case, it is more pertinent for the thin emergent droplets, because it does not rely on measuring a contact angle and it can be equally studied for both on- and off-lattice models.
When analyzed $I$ as a function of $\hat{\nu}_T$ for fixed $n_\text{max}$ or $\widehat V_0$, it shows the existence of a critical point $\hat{\nu}_{TB}$, where it vanishes following a power laws.  Again the exponent is universal and all models are consistent with the law
\begin{align}
 I &= I_0 \left( \hat{\nu}_{TB} - \hat{\nu}_T \right)^{3.0\pm 0.2}, \quad \hat{\nu}_T \leq \hat{\nu}_{TB}, \label{eq.fitI}
\end{align}
as shown in Fig.~\ref{fig.criticalI}.  The fitted exponents and the  non-universal  critical tuning rates $\hat{\nu}_{TB}$ and  amplitudes $I_0$  are given in the Table~\ref{tablaVC}. Here, no transition is observed when  $n_\text{max}=1$ for all on-lattice cases. In these cases, the value of $I$ decreases smoothly when increasing $\hat{\nu}_T$, but never vanishes  (verified up to $\hat{\nu}_T=0.6$). This behavior indicates that even for large values of $\hat{\nu}_T$ the  structure in droplets on the wall persists. This is compatible with our results obtained in Ref.~\cite{SS2017}. For the off-lattice models, in the range of parameters considered, the transition is always present.

\begin{table*}
\caption{
Critical behavior obtained for the different order parameters (first column) and models (second column). Values considered for $n_\text{max}$ or $\widehat{V}_0$ (third column). Best fit values for the universal exponent (fourth column), the non-universal critical values of $\hat{\nu}_{TA}$($\hat{\nu}_{TB}$) (fifth column) and $Z_0$ ($I_0$) (sixth column).
The values of the $\hat{\nu}_T$ column are multiplied by $10^{3}$ (mean value and error). For the crossover cases, the values of $\hat{\nu}_{TA}$($\hat{\nu}_{TB}$) and  $Z_0$($I_0$) are simply indicative.}
\label{tablaVC}
\begin{tabular}{P{1.5cm} P{1.25cm} P{1.5cm} P{4.0cm} P{2.0cm} P{2.0cm}}  
\hline
\hline
Order parameter  & Model & $n_\text{max}$($\widehat{V}_0$) &Exponent &  $\hat{\nu}_{TA}$($\hat{\nu}_{TB}$) & $Z_0$($I_0$) \\
\hline
\hline
$\langle Z\rangle$       &  1 & 1  &                            No transition & --- & ---\\
                                    &  1 & 2               &  $0.99\pm 0.05$       & $2.45\pm0.05$   &  142554   \\
                                    &  1 & 3               &  $0.99\pm 0.04$       & $1.86\pm0.05$   &  165637  \\
                                    &  1 & 4               &  $1.01\pm 0.03$       & $1.41\pm0.05$   &  179074 \\
                                    &  2 & 2               &  $1.01\pm 0.04$       & $1.83\pm 0.05$  &  171739     \\
                                    &  3 & 2                &   $0.99\pm 0.02$       & $2.50\pm 0.05$   &  143153 \\
                                    &  4 & 2                &   $1.00\pm 0.04$       & $5.50\pm 0.05$  &    31400  \\ 
                                    &  2 & 3               &  $0.99\pm 0.03$       & $1.32\pm 0.05$  &  173272   \\
                                    &  3 & 3               &  $1.01\pm 0.05$       & $1.83\pm 0.05$  &  160822 \\
                                    &  4 & 3               &  $0.98\pm 0.06$       & $4.41\pm 0.05$  &    35274  \\
                                   \hline
                                    &  5 & $0.0002$ & No transition & --- & --- \\
                                    &  5 & 2               &  $1.00\pm 0.06$       & $4.6\pm 1.0$      &    12070 \\
                                    &  6 & 2               &  $1.00\pm 0.04$       & $1.6\pm 0.4$      &      9715 \\
                                    &  7 & $0.2$  &  $1.01\pm 0.03$             & $2.2\pm 0.8$      &      6308 \\
                                    &  8 & $0.2$  &  $0.99\pm 0.05$             & $2.4\pm 0.8$      &      6705\\
                                    &  9 &  $0.0002$  &  No transition & --- & ---\\
                                    &  9 &  $2$           &    $1.02\pm 0.05 $    & $2.8\pm  0.4 $& 39\\
                                    &  10 &  $2$  &  $1.01\pm 0.05 $          & $6.6\pm  1.0 $& 35\\
\hline                                    
\hline
                      $I$        &  1 & 1 &                    No transition & --- & --- \\
                                   &  1 & 2 &                   $3.00\pm 0.06 $         & $97\pm 1$ & 54 \\
                                   &  1 & 3 &                   $2.99\pm 0.04 $         & $38\pm 1$ & 5799\\
                                   &  1 & 4 &                   $3.02\pm 0.06 $         & $26\pm 1$ & 32183\\  
                                   &  2 & 2 &                   $3.00\pm 0.06 $         & $41\pm 1$ &   4084\\
                                   &  3 & 2 &                   $3.00\pm 0.06 $         & $62\pm 1$ &   1029\\
                                   &  4 & 2 &                   $3.00\pm 0.06 $         & $108\pm 1$ & 101\\                                                    
                                   &  2 &  3 &                  $2.99\pm 0.12 $          & $29\pm  1$ & 13115\\
                                   &  3 &  3 &                  $2.95\pm 0.12 $          & $36\pm  1$ & 8072\\
                                   &  4 &  3 &                  $3.03\pm 0.23 $          & $56\pm  1$& 1285\\
                                   \hline
                                   &  5 & $0.0002$ &     $3.00\pm 0.32 $           & $196\pm  10$  &  139 \\
                                   &  5 &  $2$  &             $3.10\pm 0.27 $          & $488\pm  20$& 4.9\\
                                   &  6 &  $2$  &             $3.03\pm 0.11 $          & $498\pm   20$& 1.4\\
                                   &  7 &  $0.2$  &    $2.96\pm 0.21 $          & $416\pm  20 $& 5.3\\
                                   &  8 &  $0.2$  &    $3.12\pm 0.28 $          & $456\pm  20 $& 5.8\\
                                   &  9 &  $0.0002$  &     $3.00\pm 0.21 $     & $418\pm  20 $& 7.6\\
                                   &  9 &  $2$           &    $3.05\pm 0.31 $          & $390\pm  20 $& 8.6\\
                                   &  10 &  $2$  &  $3.02\pm 0.25 $          & $726\pm  20 $& 0.5\\
                                 
 \hline                                    
\hline
                                    
\end{tabular}
\end{table*}

Changing $\phi$, the critical turning rates $\hat\nu_{TA}$ and $\hat\nu_{TB}$ change, but the phenomenology of the transitions and the morphology of the wetting film and droplets remain.

\section{Conclusions and discussion}\label{sec.conclusion}

We have numerically studied the steady states of ten models for self-propelled active particles in presence of a wall. We found that in all cases, the system presents three wetting phases: total wetting, partial wetting and dewetting. The total wetting to partial wetting and the partial wetting to dewetting transitions are both continuous and  appropriate order parameters vanish at the transition, following power laws. Notably, although the non-universal amplitudes and critical turning rates show very different values (see Table~\ref{tablaVC}),  the associated exponents are the same for all cases. Within the type of models considered, which have short-range interactions, short temporal memory, two spatial dimensions, and no conserved or slow field other than mass density, and that  differ on the motion being either on- or off-lattice, on the lattice geometry, on the particle-particle repulsive mechanism, on the duration of the temporal memory, on the way the particle director changes, and on the form the impenetrable wall is implemented, all models appear to belong to the same universality class. It remains to be understood whether this corresponds to some other known class of non-equilibrium phase transitions or whether even it can be mapped to an equilibrium class. 

There is however, some differences between the off- and on-lattice descriptions that could signal their separation into two  universality subclasses.  In-lattice models show a  maximum in the mass autocorrelation function, which indicates the periodic arrangement of droplets, maximum that is absent  off-lattice. The analysis performed in Ref.~\cite{SS2017} showed that the position of the maximum diverges at the transition, signaling an increment in the mean distance between  droplets. This distance becomes much larger than the lattice spacing, hence it is unlikely  that the periodicity is attributed to the underlying lattice structure. 
On the other hand, in the off-lattice models, particles permanently push others in contact with the walls, inducing a motion on the droplets along the wall, effect that could destroy the periodicity. However, this can not be either the full explanation, as the periodicity is also absent in Model 10, where particles do not slide. The second difference, is that the second transition always takes place for the off-lattice cases, even for very small values of $\widehat{V}_0$, contrary to in-lattice, where the transition does not take place for $n_\text{max}=1$. 
      
Finally, we remark that the wetting phenomenon studied here is different to the equilibrium case. First, for wetting to take place in equilibrium, it is necessary that particles are attracted to the walls. In active systems, it is the persistence that makes particles wet the walls and no attraction is needed. Also, in equilibrium droplets are macroscopic objects and it is possible to determine a contact angle, which acts as an order parameter, while for the models reported in this article, droplets are made of only few particles and other order parameters should be considered.

%%%%%%%%%%%%
\acknowledgments 
This research was supported by the Millennium Nucleus ``Physics of active matter'' of the Millennium Scientific Initiative of the Ministry of Economy, Development and Tourism (Chile) and Fondecyt Grants No.\ 1151029 (N.S.) and 1180791 (R.S.).

%%%%%%%%%%%%


\begin{thebibliography}{99}
\bibitem{commentalpha}
Strictly speaking, for a tumbling rate $\alpha$,  the tumbling probability at each time step is $(1-e^{-\alpha})$. For small values, as we consider in this article, it is well approximated by $\alpha$.
\end{thebibliography}
\end{document}